\documentclass[12pt, draftcls, journal, compsoc, onecolumn]{IEEEtran}
\usepackage[pdftex]{graphicx}
\usepackage{amsmath}
\usepackage{amsfonts}

\usepackage{url}
\usepackage{array}
\title{Computer-aided modelling of complex physical systems with BondGraphTools}
\usepackage{setspace}
\usepackage{subcaption}
\usepackage{hyphenat}
\usepackage{tabularx}
\usepackage{xcolor}
\graphicspath{{images/}}

\definecolor{codebg}{HTML}{F6F6F6}
\definecolor{codeframe}{HTML}{CCCCCC}

\usepackage{listings}
\newcommand{\D}[2]{\frac{\mathrm{d} #1}{\mathrm{d} #2}}

\newcommand{\df}[1]{\mspace{2mu}  \mathrm{d}#1}
\renewcommand{\P}[2]{\frac{\partial #1}{\partial #2}}

\newcommand{\BGT}{\texttt{BondGraphTools}}

\author{
\IEEEauthorblockN{	
	Peter~Cudmore\IEEEauthorrefmark{1}\IEEEauthorrefmark{3},
	Peter~J.~Gawthrop\IEEEauthorrefmark{1},~
	Michael~Pan\IEEEauthorrefmark{1},
	Edmund~J.~Crampin\IEEEauthorrefmark{1}\IEEEauthorrefmark{2}\IEEEauthorrefmark{3}}\\
\IEEEauthorblockA{
	\IEEEauthorrefmark{1}
	Systems Biology Laboratory, School of Mathematics and Statistics,\\ and Department of Biomedical Engineering,\\ University of Melbourne, Parkville, Victoria 3010
}\\
\IEEEauthorblockA{
	\IEEEauthorrefmark{2}
	School of Medicine, Faculty of Medicine, Dentistry and Health Sciences,\\ University of Melbourne, Parkville, Victoria 3010}\\
\IEEEauthorblockA{
\IEEEauthorrefmark{3}
ARC Centre of Excellence in Convergent Bio-Nano Science and Technology,\\ Melbourne school of Engineering, University of Melbourne, Parkville, Victoria 3010
}
}
\begin{document}
	\maketitle
	\begin{abstract}
		BondGraphTools is a Python library for scripted modelling of complex multi-physics systems.
		In contrast to existing modelling solutions, BondGraphTools is based upon the well established bond graph methodology, provides a programming interface for symbolic model composition, and is intended to be used in conjunction with the existing scientific Python toolchain.
		Here we discuss the design, implementation and use of BondGraphTools, demonstrate how it can be used to accelerate systems modelling with an example from optomechanics, and comment on current and future applications in cross-domain modelling, particularly in systems biology.
	\end{abstract}

\section{Introduction}
From modelling mitochondrial electron transport~\cite{Gawthrop2017aa} to predicting the future of ecological systems~\cite{Cellier2006} and controlling nano-electro-mechanical oscillator arrays~\cite{Holmes2012aa}, the dynamics of complex multi-physics systems is of great interest to scientists, engineers and mathematicians.
Often complex systems can be represented by many distinct processes interacting via a network topology and understanding such systems is currently an area of active interdisciplinary research. 
Computational tools are crucial for understanding complex systems, in particular for performing simulations, parameter fitting and sensitivity analysis as such systems are often too large and heterogeneous for analytic approaches to be tractable. 
In some cases, software is also necessary to manage the scale of such problems, as researchers from many different laboratories can be working on sub-models of the same problem simultaneously.
For large scale collaborative tools to function, there must be an agreement on how sub-models can be assembled into a whole, an interface and supporting computational infrastructure.

Complex \emph{physical} systems have some features that complex systems more generally do not; they must obey the laws of physics. 
Consider the contrast between resource flows in an ecological system and a network model of social interactions. 
While one could certainly ignore conservation laws when modelling an ecosystem, the quantities of interest are physical things; resources, temperature and energy more generally, all of which are locally conserved. 
In the case of social networks one is concerned with the flow of information, patterns and signals, none of which appear to obey the same kind of conservation laws one expects from physical systems.
 
It is the laws of physics that provide a justifiable interface between different models of physical subsystems. 
In particular any \emph{connection}, that is, any non-physical feature of the modelling framework set up to allow disparate sub-models to interface, between two models of physical processes must conserve energy. 
Bond graphs are an example of such a modelling framework.

\begin{figure}
	\caption{The transfer of power through a bond between port $j$ on system A and port $k$ on system B..
		\label{fig:power_bond}}
	\centering
	\includegraphics{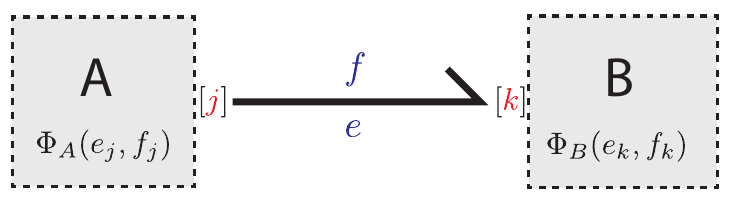}
\end{figure}

Invented by Paynter in the 1960's~\cite{Paynter1961a}, bond graphs are a port-based modelling tool that describe the flow of power through a network of energy storage and dissipation sub-systems.
Ports attached to a particular subsystem can be related via an energy conserving 'power bond' and are defined in terms of \emph{force-like} efforts $e$ and \emph{flux-like} flows $f$ such that the (signed) power transfer at any instant is $P=ef$. 
Figure~\ref{fig:power_bond} shows an example of a power bond. 
Here, the power entering system A is given by $P_{\text{A};j} = e_jf_j$, similarly for B; $P_{\text{B};k} = e_kf_k$, such that power is conserved between systems, i.e. $P_{\text{A};j}+P_{\text{B};k} = 0$. 
The directed power transfer through the bond is therefore 
\begin{equation}
P_{bond}=ef,\ \text{with}\ e = e_j = e_k \ \text{and}\ f = -f_j = f_k.
\label{eq:bond}
\end{equation}
The behaviour of systems A and B are captured via $\Phi_\text{A}$ and $\Phi_\text{B}$ respectively.
We refer interested readers to the tutorial by Gawthrop and Bevan~\cite{Gawthrop2007aa} for an overview of bond graph modelling and for an in depth treatment see~\cite{KarMarRos12}.

One interpretation of bond graphs is as formalising modelling via analogy. In bond graph terms, voltage and pressure are both effort, while current and mass flow are flows hence giving an explanation for the hydraulic analogy of electricity. 
More recently, there has been an effort spearheaded by van der Schaft~\cite{VanderSchaft2014b} and others, to integrate port-based bond graphs approaches with modern geometric mechanics resulting in theory of port-Hamiltonian systems. This follows from the observation that for conservative systems effort and flow variables can be related to the time derivatives of an energy storage function. 

It seems to be that, regardless of modelling methodology, as model complexity increases software becomes increasingly necessary. Broadly speaking, there are two main computational approaches to cross-domain systems modelling; computer aided design tools and mathematical software, both of which have some scope for bond graph modelling.

The first are Computer Aided Design (CAD) tools which follow in the tradition of electrical design automation and include Dymola\footnote{\url{www.3ds.com/products-services/catia/products/dymola/}} and 20-sim\footnote{\url{www.20sim.com}} in which users draft technical schematics of the system from a library of components. The other main approach is via specialised mathematical software such as MATLAB (Simulink), Maple (MapleSim) or Mathematica (SystemModeller) in which again users construct graphical representations of the model.  Almost all of the existing software implementations of bond graph modelling are graphical in nature and within a proprietary or isolated software ecosystem. For some applications, this is beneficial; vendors can provide a standardised visual interface with integrated analysis tools. For other modelling problems, the existing software lacks the capacity for automation, is hampered by restricted access to source code and is difficult (if not impossible) to integrate with existing algorithms.

Hence, there is need for a symbolic physical systems modelling toolkit that is open-source, easy to integrate into existing workflows, and written in an accessible and widely used scripting language.
In particular, there is a great need for infrastructure to support automated model building and simplification.


Here we introduce \BGT, a python library for building and manipulating symbolic models of complex physical systems, built upon the standard scientific python libraries.
The \BGT\ package is different from existing software in both design and implementation in that it:
\begin{enumerate}[\IEEEsetlabelwidth{6}]
\item is explicitly based on physical modelling principles.
\item provides an \emph{application program interface} (API) as opposed to a graphical user interface (GUI).
\item is designed for \emph{symbolic} model composition and order reduction, as opposed to being primarily a numerical toolkit.
\item is intended to be used in conjunction with the standard python libraries, as opposed to being used as a stand-alone software package.
\item allows for modification and integration as it is open source, version controlled, and readily available, instead of closed-source and proprietary.
\item is designed with modularity and extensibility in mind, as opposed to being a monolithic software suite.
\end{enumerate}

As a programming interface, \BGT\ gives modellers a means to integrate the tools and techniques of software development into their modelling workflow.
This includes being able to script tasks like model re-parametrisation and batch replacement of model subcomponents both of which can be tedious and time-consuming in graphical environments. 

Building upon and integrating with the existing python ecosystem means that \BGT\ can specialise in providing an interface for model building without concerning itself with other tasks. This results in a smaller codebase, and hence more sustainable software.
Using modern open source practices allows other developers to easily modify, contribute and build upon the \BGT\ codebase. Unlike proprietary software, users are free to implement new features and extend \BGT\ as they see fit, for example by using \BGT\ as a foundation for graphical modelling environments.

For a large class of systems, particularly in systems biology, only the network topology of a system may be known at the time of modelling.
As \BGT\ represents model parameters symbolically, values are free to be determined later in the modelling process via existing parameter estimation techniques.

\BGT\ provides a programming language for building bond graph models which are automatically turned into differential equations to be analysed or simulated.
Python is well established as a robust and easy-to-use general purpose programming language with a wide variety of standardised and well supported libraries for standard scientific tasks, and \BGT\ adheres to python language idioms by emphasising self-explanatory, self-documenting and self-contained code.

In section~\ref{sec:basicuse} we provide an overview of the basic interface for building models in \BGT\ with python code examples.
The model reduction and simplification approach is detailed in section~\ref{sec:reductions} .
Section~\ref{sec:example} demonstrates how \BGT\ can effectively scale up systems modelling and simulation by considering an recent example from quantum optomechanics.
We outline the sustainable software development practices used to ensure the code quality, robustness and availability of \BGT\ in section~\ref{sec:discussion} and close with some comments about direction of \BGT's continued development.

\section{Design Motivation and Basic Use}\label{sec:basicuse}
In order to model complex physical systems, we must have an idea of what the model subsystems are, and how they can be related.
Here we outline the relevant classes of models and how they are related, both abstractly and in practical usage.
Throughout the code examples, we assume \BGT\ has been imported as follows
\begin{lstlisting}
import BondGraphTools as bgt
\end{lstlisting}
We demonstrate code as it appears in a python script, which could be equally executed by entering the code into an IPython session or a Jupyter notebook~\cite{Perez2007}.

In \BGT, and bond graph modelling in general, all models have:
\begin{itemize}
	\item at most one parent model, and any number of child models,
	\item a (possibly zero) number of power \emph{ports} each of which has two associated variables, an effort-like $e$ (voltage, force, etc.) and a flow-like $f$ (current, velocity, etc.) such that power $P=ef$ is positive when the process is consuming or accumulating energy through that particular port,
	\item a set of acausal equations, or constitutive relations, that characterise the behaviour for a particular model and are either derived from child models or specified a-priori; for example generalised linear dissipation (friction, Ohm's law) which relates the effort and flow via the implicit relation $0 = e - Rf$. We note that the power entering this model $P = ef = Rf^2$ is positive semi-definite for $R>0$ indicating, as one would expect, that resistance always consumes power.  
	\item a (possibly zero) number of parameters/controls which govern the behaviour of that particular model. One would consider $R$ in the above example to be a parameter if $R$ is constant, and a control in all other cases.
	\item a (possibly zero) number of state variables (with associated derivatives), related to power ports via the model constitutive relations. For example, a linear potential energy storage has governing equations 
	\[ Ce - x = 0,\qquad f - \dot{x} =0,\quad \implies P = \frac{1}{C}x\dot{x} \]
	so that the energy $E(t)$ stored in the state variable $x(t)$ at a particular time $t$ in that component (up to a constant) is $E(t) =\int_{0}^tP\df{t} = \frac{x(t)^2}{2C}$. 
\end{itemize}

Thus model building in \BGT\ consists of instantiating the models one wishes to use, defining parent-child relationships, then specifying energy sharing between ports.
Once a model has been constructed, simplified equations are automatically derived using symbolic algebraic tools.
Having a simplified symbolic representation of the system is valuable as it provides modellers a way to export the equations into whatever format they desire. 
In particular, the set of implicit equations can easily be fed into standard parameter estimation routines, discretised for implementation in other architectures, or passed to solver routines.

\subsection{Creating models}
\label{sec:creation}
Models in \BGT\ are broadly split into two classes: \emph{composites}, which are assembled using other models in a has-a relationship; and \emph{atomics}, which represent processes that are considered indecomposable and often fundamental, both of which are constructed using the \lstinline{new} function. 

New composites can be constructed using 
\begin{lstlisting}
model = bgt.new(name="New Model")
\end{lstlisting}  
which results in the variable \lstinline{model} containing a new instance of the \lstinline{BondGraph} class, the composite base class, with no components and with the name `New Model'.
This is identical to creating new instances of the \lstinline{BondGraph} class directly via
\begin{lstlisting}
model = bgt.BondGraph(name="New Model")
\end{lstlisting}
which is available for the purposes of providing an object oriented interface, of which an example can be found in \ref{sec:examplelinearosc}.

New atomics can be created in a similar manner by specifying the model class and value
\begin{lstlisting}
dissipation = bgt.new("R", value=1)
\end{lstlisting}
so that each variable now contains new instances of the respective atomics. This particular atomic has only one parameter, the dissipation rate, which can be unambiguously assigned the value of $1$. A list of common models can be found in Table~\ref{tab:equivalent_models}.
In keeping with the emphasis on symbolic equations, parameters can be either numeric or symbolic values.


\begin{table}[!t]
	\footnotesize
	\caption{\label{tab:equivalent_models}Equivalent models in \BGT.}
	\renewcommand{\arraystretch}{1.3}
	\begin{tabular}{l | l}
		Model & Physical Process\\
		\hline
		R & Dissipation, friction, Ohms law.\\
		C & Potential energy, compliance, capacitance\\
		I & Kinetic Energy, inertance, inductance.\\ 
		TF & Transformer, lever, gearbox\\
		GY & Gyrator, DC motor\\
		Se & External force, voltage source\\
		Sf & Current source, rigid connection \\
		0 & Kirchhoff's current law, conservation of mass\\
		1 & Kirchhoff's voltage law, Newton's first law
	\end{tabular}
\end{table}

Two models of particular interest are junction laws and port-Hamiltonians. In bond graphs, network conservation laws, such as Kirchhoffs Laws, are themselves considered atomics, and hence must be added just as any other model.
For example, the 0 junction describes the conservation law where all efforts are equal. Suppose there are $n$ ports associated with this junction, then the constitutive relation is
\[
0 = e_k - e_0\quad \forall\ k\in 1, \ldots, n-1, \qquad \sum_{k=0}^{n-1} f_k = 0.
\]
It is clear by inspection that this is indeed power conserving. Similarly, the $n$-port 1 junction, or `common flow', has relations given by
\begin{equation}
0 = \sigma_jf_j - \sigma_0f_0\quad \forall k \in 1, \ldots{n-1},\qquad \sum_{k=0}^{n-1}\sigma_ke_k =0 \label{eq:one_junction}
\end{equation}
where $\sigma_k = 1$ if the associated port $k$ is oriented inwards, or $\sigma_k = -1$ if oriented outwards. Explicitly associating the direction with the port (and hence the junction), instead of determining it circumstantially from the bond means that the governing equations are independent of any particular choice of bond orientation.

\begin{figure}[!t]
	\centering
	\caption{\label{fig:Hamiltonian_BondGraph}Bond graph of a Hamiltonian system.}
	\includegraphics{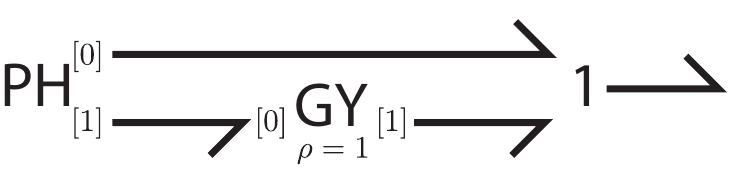}
\end{figure}

Port-Hamiltonians on the other hand are interesting as they present an easy, and physically relevant, way to introduce nonlinearity and non-trivial coupling between subsystems. If we understand $e=(e_1,e_2,\ldots)$ and $f = (f_1,f_2,\ldots)^T$ to be vectors, then port-Hamiltonians have constitutive relations generated by
\begin{equation}
0 = e - \nabla_x H(x), \qquad  0 = f - \dot{x}. \label{portHamiltonEq}
\end{equation}
Here, the port-Hamiltonian energy function $H$ is not encoded a-priori with a symplectic structure as per traditional Hamiltonian mechanics.
In Hamiltonian mechanics, the state of a given object $k$ is encoded in position-momentum pairs $(q_k, p_k)$ such that the total energy in the system is given by a function $H$. If a system is closed, energy is conserved and thus $H$ is constant. It is a well known fact that in such systems the state $x=(q,p)$ evolves according to Hamilton's equations
\begin{equation}
\D{x}{t} = S\ \nabla_xH(x), \quad S =\left(\begin{matrix}0 & I\\-I & 0\end{matrix}\right). \label{eq:Hamiltons}
\end{equation}
Port-Hamiltonian theory encodes the skew symmetric matrix $S$ (and hence symplectic geometry) via network laws~\cite{VanderSchaft2014b} using symplectic gyrators~\cite{Maschke1992aa}. A gyrator~\cite{Tel1948} is a power conserving two-port component with the constitutive relations
\begin{equation}
\left(\begin{matrix} 
e_0 - \rho f_1\\
e_1 + \rho f_0
\end{matrix}\right) = 0 \label{eq:gyrator}
\end{equation}
parametrised by the \emph{gyration resistance} $\rho$, and hence a symplectic gyrator can be generated by setting $\rho = 1$.
Figure~\ref{fig:Hamiltonian_BondGraph} demonstrates an example of how this can be built with a bond graph.
Here, we take $x_0 = q$ and $x_1 = p$ so that the port-Hamiltonian constitutive relations are equivalent to 
\[
(e_0, e_1, f_0, f_1)_\text{PH} = \left(\P{}{q} H, \P{}{p} H, \dot{q},\dot{p}\right)
\] 
for a given Hamiltonian $H(q,p)$. As per~\eqref{eq:bond}, the bond between the port-Hamiltonian and gyrator gives
\[
\P{}{p}H = e_{\text{PH};1} = e_{\text{GY};0} = f_{\text{GY};1}
\]
and by treating signs carefully we have
\[
\dot{p} = f_{\text{PH};1} = -f_{\text{GY};0} = e_{\text{GY};1}.
\]
Finally, it follows from \eqref{eq:one_junction} that
\[
e_{\text{GY};1} + e_{\text{PH};0} = e_{\text{1;out}} \quad \implies\quad  \dot{p} + \P{H}{q} = e_{\text{1;out}}
\]
and that 
\[
0 = f_{\text{GY};1} - f_{\text{PH};0} \implies \dot{q}  - \P{H}{p} = 0.
\]
For closed systems, where $e_{\text{1;out}} = 0$, this is nothing but Hamilton's equations~\eqref{eq:Hamiltons} for a two dimensional system.
Incorporating port-based modelling with Hamiltonian mechanics provides an important bridge between physics and engineering which will be seen in section \ref{sec:examplecavity}.

\subsection{Defining relationships}
There are two categories of relationships between models in \BGT; structural and energetic.
Structural relationships describe composition, how a given model can contain many simpler models. This organises systems into a tree-like hierarchy of 
parent-children relations between models.
Energetic relationships describe how the ports of a given set of child models are connected within a given parent model.

\subsubsection{Structural relationship}
\label{sec:heirarchy}
In \BGT, permissible models (that is, models  which have no established relationships) can be added (removed) from composite models with the \lstinline{add} and \lstinline{remove} functions. For example, adding a dissipation component to the composite model can be achieved with
\begin{lstlisting}
bgt.add(model, dissipation)
\end{lstlisting}
where the first argument is the parent model, and the remaining arguments are the intended children or components. 

A file system interface navigates the model hierarchy and is implemented within the \lstinline{ModelBase} class from which all models inherit.
In particular,
\begin{itemize}
	\item \lstinline{ModelBase.uri} is a uniform resource identifier (URI) locating that particular model.
	\item \lstinline{ModelBase.parent} refers to the parent model if it exists. 
	\item \lstinline{ModelBase.root} refers to the top of the model tree.
	\item For composite models, \lstinline{Composite.components} will contain a list of sub-models.
\end{itemize}
Atomics are thus analogous to files, and composites to directories, with the root model analogous to the unix hostname. 

\subsubsection{Energetic Relationships}
\label{sec:bonds}
The energetic relationship defines how power is transferred between components inside a particular model via ports which belong to components, and depends on first establishing a model hierarchy.
Once a set of component-wise relationships is established within a model, components are connected to each other (and hence energy bonds defined, in the bond graph terminology).
When it is not ambiguous, for example when connecting a one-port component to a junction with arbitrarily many identical ports, it is sufficient to connect the component directly via
\begin{lstlisting}
bgt.connect(component_1, component_2)
\end{lstlisting}

Otherwise one must specify the port by index
\begin{lstlisting}
target_port = (component_2, 0)
\end{lstlisting}
and connect the port directly:
\begin{lstlisting}
bgt.connect(component_1, target_port)
\end{lstlisting} 

Composite models keep track of the energy relationships (bonds) in the member attribute \lstinline{Composite.bonds}. 
Similarly, two components can be disconnected via the \lstinline{disconnect} method which has an identical interface to \lstinline{connect}.

\subsection{Model Attributes}
Composite and Atomic models often have associated parameters which can be accessed by the member attribute \lstinline{ModelBase.params}.
Once the structure of a composite model has been established, and the internal connections defined, one can generate governing equations for the entire model via \lstinline{ModelBase.constitutive_relations} in terms of the derived state variables \lstinline{ModelBase.state_vars} to be used for further analysis.

\section{Symbolic Composition and Reduction}
\label{sec:reductions}
Core to \BGT\ is symbolic model reduction. 
This relies on the fact that each model or component has a set of constitutive relations $\Phi(X) = 0$ which are implicit equations defining the model behaviour in terms of how power is manipulated.
\subsection{Model structure}
The local co-ordinate space of a model $\alpha$ is taken to be $\mathcal{X}_\alpha = \left\{\dot{x}_\alpha, e_\alpha, f_\alpha, x_\alpha, u_\alpha \right\}$ where $x, \dot{x}$ are the vectors of storage co-ordinates and time derivatives respectively, $e,f$ are the power interconnection variables effort and flow, and $u$ are controls or inputs.
Let us assume that 
$\dim{x_\alpha} = n_\alpha$, $\dim{e_\alpha} = \dim{f_\alpha} = m_\alpha$, $\dim{u_\alpha} = k_\alpha$ for finite $n_\alpha,m_\alpha,k_\alpha$ and $n_\alpha + m_\alpha \ge 1$ so that there is non-trivial behaviour.
We also define $N_\alpha = 2(n_\alpha + m_\alpha) +  k_\alpha$.
One can define a matrix $L_\alpha\in \mathbb{R}^{N_\alpha \times N_\alpha}$, a vector field $V_\alpha:\mathcal{X_\alpha}\rightarrow \mathbb{R}^{N_\alpha}$ such that the constitutive relations are equivalent to
\[
\mathbf{0} = L_\alpha X + V_\alpha (X) \qquad \forall X \in \mathcal{X}_\alpha.
\]
We expect $L_\alpha$ to be sparse and $\mathop{\mathrm{rank}}L_\alpha \le n_\alpha + m_\alpha +  k_\alpha < N_\alpha$ so that there is at least one eigenspace of $L_\alpha$ per state variable pair, unconnected port or control input.
\subsection{Composition}
Any number of constitutive relations $\Phi = [\Phi_\alpha, \Phi_\beta,\ldots] = 0$ can thus be combined via
\begin{align}
0 &= LX+V(X) \notag\\ &=  \left(
\begin{matrix}
L_\alpha & 0 &\dots \\
0 & L_\beta & 0 \\ 
\vdots&0 & \ddots
\end{matrix}
\right) X + \left(\begin{matrix}V_\alpha\circ \pi_\alpha \\ V_\beta\circ \pi_\beta \\\vdots \end{matrix}\right)(X) 
\label{eq:implicit_eq}
\end{align}
where $X\in \mathcal{X} =  \mathcal{X}_\alpha \oplus \mathcal{X}_\beta \oplus \ldots$ is the direct sum of local co-ordinate spaces and 
$\pi_\alpha:\mathcal{X} \rightarrow {X}_\alpha$ is a projection back into local co-ordinates.
One can easily incorporate interconnecting power bonds by noting that a bond connecting port $i$ on component $\alpha$ to port $j$ on component $\beta$ can be represented by the rows
\[
0 = e_{\alpha;i} - e_{\beta;j}  = \left[\theta_{\alpha;i}, -\theta_{\beta;j} \right]\left(\begin{matrix}X_\alpha\\ X_\beta\end{matrix}\right)\]
and\[
0 = f_{\alpha;i} + f_{\beta;j}  = \left[\theta_{\alpha;i}, \theta_{\beta;j} \right]\left(\begin{matrix}X_\alpha\\ X_\beta\end{matrix}\right)\]
where $\theta_{\alpha;i}$ is the co-basis vector such that $\theta_{\alpha;i}X_\alpha = e_{\alpha;i}$ i.e,.; $\theta_{\alpha;i}$ is a row vector with one in the column corresponding to $e_{\alpha;i}$ (similarly for $\theta_{\beta;j}$).
It follows that the set of all bonds form a junction structure on the larger space $X$ represented by a full rank matrix $J$ such that $JX = 0$ with row rank identical to the number of bonds.
It is convenient to simply consider this as an additional constitutive relation, and append it to the linearised matrix $L$ in \eqref{eq:implicit_eq}.

Given the space $X$, there exists an orthonormal permutation matrix $P$ such that $X' = P^{-1}X$, and \[X' =(\dot{x}_\alpha, \dot{x}_\beta,\ldots e_\alpha, f_\alpha, e_\beta, f_\beta, \ldots, x_\alpha, x_\beta,\ldots, u_\alpha, u_\beta,\ldots).\] 
It also follows from elementary linear algebra that there is exists an invertible matrix $\Lambda$ such that
$L' = \Lambda L P$ is in a reduced upper triangular form (reduced row echelon form with leading terms always on the diagonal), resulting in an exact simplification of \eqref{eq:implicit_eq}
\begin{equation}
\Lambda L P X + \Lambda V(X) = L'X' + V'(X') = 0 \label{eq:simplified}
\end{equation}
where $V'(X') = \Lambda V(PX')$.
The coordinate ordering, and hence the permutation matrix $P$ is chosen so that (in the linear case) triangularisation produces the correct order of dependence when substitution is performed; rates of change, efforts and flows are expressed in terms of state and control variables. 
This can be trivially extended to simple nonlinear cases as non-zero diagonal entries of $L'$ determine substitution rules, though more complicated systems can however produce irreducible algebraic constraints.
\subsection{Output}
Constitutive relations, retrieved via member attribute \lstinline{ModelBase.constitutive_relations} by evaluating ~\eqref{eq:simplified}, can be generated for any model at any level of the structural tree, and is performed in a recursive manner to reduce the number of calculations. 
Reduced symbolic models can be passed into a simulation service, which renders~\eqref{eq:simplified} as a function, then initialises and solves the associated initial value problem, for example by

\noindent
\begin{minipage}{\linewidth}
\begin{lstlisting}
u = "sin(t)"                     
t,x = simulate(model, 			 
			   x0=x0,            
			   timespan=[0,100], 
			   control_vars=[u]),
			   dt=0.1            
\end{lstlisting}
\end{minipage}

In the current release, \BGT\ uses \texttt{DifferentialEquations.jl}~\cite{Rac2017} to provide numerical solvers, as it is provides a simple and fast interface to the SUNDIALS~\cite{sundials} set of differential-algebraic equation (DAE) solvers.
This particular method was chosen based on a trade-off between development convenience and utility; DAE solvers are necessary as the constitutive relations of models can involve algebraic constraints that cannot be eliminated. \BGT\ dynamically generates Julia~\cite{Julia} code, uses the \texttt{DifferentialEqations.jl} interface to solve the DAE, and passes the results back to python for analysis and plotting.

An important feature is the ability to specify control functions as a string. This allows users to quickly test different control schemes for any particular system, and supports most common mathematical functions.

\section{Example: an optomechanical experiment}
\label{sec:example}
\begin{figure}[t!]
	\includegraphics[width=0.95\linewidth]{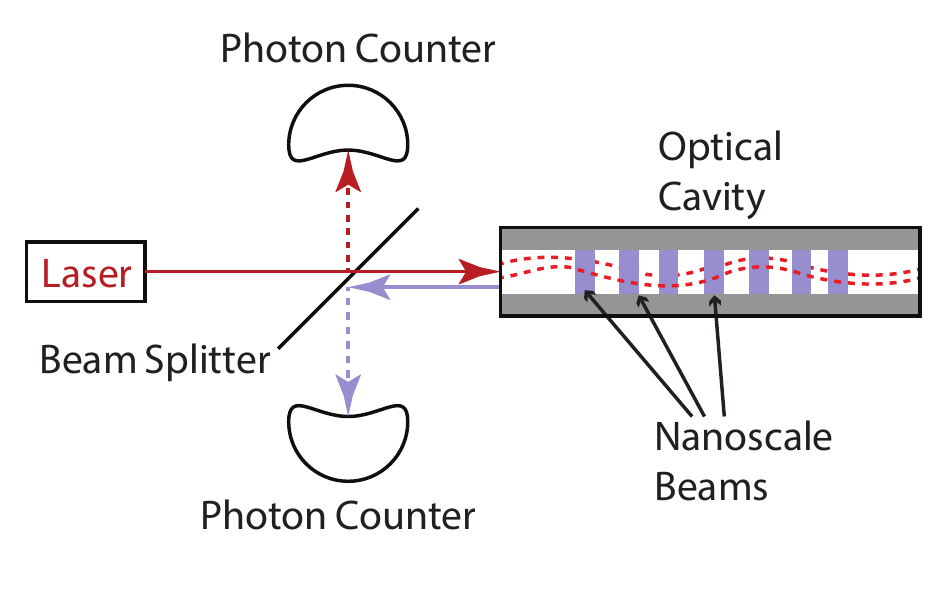}
	\caption{Illustration of the set up of an optomechanical experiment. Here an optical cavity containing many nanomechanical resonators is driven by a coherent photon source. 
		The amount of optical power that enters the cavity depends on the detuning from resonance of the cavity, which itself depends on the displacement of the mechanical resonators.
		Feedback between the subsystems occur as radiation pressure couples the mechanical modes to the optical field.}
	\label{fig:exp}
\end{figure}

As an example, we consider a coupled oscillator system from optomechanics, the experimental conditions of which are illustrated in figure~\ref{fig:exp}. 

Here, a coherent photon source is used to excite an optical cavity containing a number of nanomechanical beams.
The excitation leads to a electromagnetic standing wave within the cavity, which in turn exerts radiation pressure upon the mechanical beams, causing them to vibrate. The vibrations close a feedback loop between the cavity and beams, modulating the resonant frequency of the cavity and hence changing the amount of radiation pressure coupling~\cite{Holmes2012aa}. When there are several nanomechanical beams, the optical subsystem acts as a nonlinear coupling process for the mechanical subsystem which, under certain conditions~\cite{PC2015}, can cause the beam vibrations to synchronise even when the natural resonances are quite different. Such systems have been proposed as the basis of new quantum sensing technology, however for the purposes of this work, 
the semi-classical approximation (where quantum effects are assumed negligible) provides an example of a multi-domain system (optical, mechanical and electrical) with which to use of \BGT\ to model the system.

In this section we will define classes and functions that build the cavity model and run the experimental protocol.
This demonstrates that modelling the experimental apparatus in a modular fashion allows automation of post-processing to, for instance, map the simulation data into the same 
structure as the expected experimental output. 
\subsection{Mechanical Subsystem}
\label{sec:examplelinearosc}
In terms of model building, we first must have a way of generating many beams from a template.
\BGT\ provides both method based and object oriented functions for this task.

The class definition for linear oscillators is given by

\noindent
\begin{minipage}{\linewidth}
\begin{lstlisting}
class Linear_Osc(bgt.BondGraph):  
    damping_rate = 0.1 

	def __init__(self, freq, index):

		# Create the components
		r = bgt.new("R", name="R", value=self.damping_rate)
		l = bgt.new("I", name="L", value=1/freq)
		c = bgt.new("C", name="C", value=1/freq)
		port = bgt.new("SS")
		conservation_law = bgt.new("1")
	
		# Create the composite model and add the components
		super().__init__(
			name=f"Osc_{index}",
			components=(r, l, c, port, conservation_law)
		)
	
		# Define energy bonds
		for component in (r,l,c):
			bgt.connect(conservation_law, component)

		bgt.connect(port, conservation_law)
	
		# Expose the external port
		bgt.expose(port, label="P_in")
\end{lstlisting}
\end{minipage}

The \lstinline{__init__} function is called during the object creation process and creates a number of new components, adds them to the newly created parent model, connects the child components and exposes a port. 
Once this sequence has been completed for a particular instance, it can be used like any other model.
Once the class is defined in this manner, one can easily create new instances, for example by:
\begin{lstlisting}
oscillator = Linear_Osc(freq=1, index=0)
\end{lstlisting}

\subsection{Optical Subsystem}
\label{sec:examplecavity}
The optical subsystem is built with the method below and which will be described in this subsection.

\noindent
\begin{minipage}{\linewidth}
\begin{lstlisting}
def coupled_cavity():
	model = bgt.new(name="Cavity Model")

	# Define the interaction Hamiltonain
	coupling_args = {
		"hamiltonian":"(w + G*x_0)*(x_1^2 + x_2^2)/2",
		"params": {
			"G": 1,   # Coupling constant
			"w": 6    # Cavity resonant freq.
		}
	}
	port_hamiltonian = bgt.new("PH", value=coupling_args)

	# Define the symplectic junction structure  
	symplectic_gyrator = bgt.new("GY", value=1) 
	em_field = bgt.new("1") 
	bgt.add(model, port_hamiltonian, 
	 			   symplectic_gyrator, 
	 			   em_field)   
	bgt.connect(em_field, (port_hamiltonian, 1))
	bgt.connect(em_field, (symplectic_gyrator, 1))

	bgt.connect(
		(port_hamiltonian, 2), (symplectic_gyrator, 0)
	)
	
	# Construct the open part of the system
	dissipation = bgt.new("R", value=1)
	photon_source = bgt.new('SS')
	bgt.add(model, dissipation, photon_source)
	bgt.connect(em_field, dissipation)
	bgt.connect(photon_source, em_field)
	bgt.expose(photon_source)
	
	# Build the oscillator array
	frequencies = [2 + f for f in (-0.3, -0.1, 0, 0.1, 0.3)]
	osc_mean_field = bgt.new("0")
	bgt.add(model, osc_mean_field)
	bgt.connect(osc_mean_field, (port_hamiltonian, 0))
	osc_array = [Linear_Osc(freq, index) 
				 for index, freq in enumerate(frequencies)]

	for osc in osc_array:
		bgt.add(model, osc)
		bgt.connect(osc_mean_field, (osc, "P_in"))

	return model
\end{lstlisting}
\end{minipage}

One significant advantage of the port-Hamiltonians in general, and \BGT\ 's implementation in particular, is that one is able to 
essentially mix-and-match modelling based on ones understanding of the system in question.
In this case, the unforced, conservative optomechanical system is known~\cite{Holmes2012aa} to have a Hamiltonian of the form
\begin{equation}
H_\text{opt} = \frac{1}{2}(\omega + Gq_1) \left(p_2^2 + q_2^2 \right)  \label{eq:Ph}
\end{equation}
so that the natural frequency $\omega$ of the cavity is shifted proportional to the oscillator displacement by a factor of $G$. 
Here $q_2,p_2$ represent the average field position and momentum respectively.
Port-Hamiltonians can be created within \BGT\ by specifying the Hamiltonian (as a string), and optionally specifying parameter values.
As mentioned in section~\ref{sec:creation}, the port-Hamiltonian framework has no implicit symplectic structure; it must be explicitly defined using a symplectic gyrator in conjunction with a common flow junction.

After interactions with the external environment have been specified  the oscillator array is added.
Each oscillator is built from the class definition as per \ref{sec:examplelinearosc}, and connected up to the common-force mean field junction.
One now has a reusable function that models the given coupled oscillator system.

Here the utility of a scripting interface is particularly clear. 
Instead of adding each individual oscillator and specifying that individuals frequency as one might do in a more traditional graphical modelling environment, the oscillator array is batch-constructed from a list of frequencies.
Again, in more traditional CAD environments, each newly added a parametrised component must then be manually connected, whereas here connections are defined for every oscillator programmatically.
One could, for example, instead generate a random vector of arbitrary length which would allow a user to compare different frequency distributions and oscillator counts quickly and easily.

Once a model is built, equations of motions can be algorithmically generated. In this example, building the model and printing out the reduced equations is achieved via
\begin{lstlisting}
cavity = coupled_cavity()
print(cavity.constitutive_relations)
\end{lstlisting}
which results in the following list:
\begin{lstlisting}
[dx_0 + 23*x_11/10 + 17*x_3/10 + 19*x_5/10 + 2*x_7 
+ 21*x_9/10, 
dx_1 - x_0*x_2 - 6*x_2, 
dx_2 - e_0 + x_0*x_1 + x_0*x_2 + 6*x_1 + 6*x_2, 
dx_3 - x_1**2/2 - x_2**2/2 + 17*x_3/100 + 17*x_4/10, 
dx_4 - 17*x_3/10, 
dx_5 - x_1**2/2 - x_2**2/2 + 19*x_5/100 + 19*x_6/10, 
dx_6 - 19*x_5/10, 
dx_7 - x_1**2/2 - x_2**2/2 + x_7/5 + 2*x_8, 
dx_8 - 2*x_7, 
dx_9 - x_1**2/2 + 21*x_10/10 - x_2**2/2 + 21*x_9/100, 
dx_10 - 21*x_9/10, 
dx_11 - x_1**2/2 + 23*x_11/100 + 23*x_12/10 - x_2**2/2, 
dx_12 - 23*x_11/10, 
f_0 - x_0*x_2 - 6*x_2]
\end{lstlisting}
Here \lstinline{dx_1} is the time derivative of the state variable \lstinline{x_1}, and \lstinline{(e_0, f_0)} are the externally imposed effort and flow respectively. The result is stored in symbolic python variables, as opposed to strings, which can then be passed on to other analysis tools. 
Using the reduced equations, basic simulations can also be run from within \BGT\ and additionally the cavity model itself can be used as a subcomponent in yet another model, in the same way as the linear oscillators were used to build the system model.

\section{Discussion and Conclusions}
\label{sec:discussion}
\subsection{Model Abstraction}
The complexity of large projects, modelling or otherwise, is often made tractable by introducing abstractions which allow individuals to 'divide-and-conquer' the problem.
For this to work, one requires a way for subcomponents to interface with each other. Bond graphs bring this to modelling by framing the interconnection in terms of power, which in turn allows \BGT\ to 
define a consistent way of joining sub-models together (via ports). Hence, one avoids some connectivity issues such as failing to account for loading, whilst maintaining a sense of modularity.
Connections between ports representing different physical domains are made using the energy-transmitting TF (transformer) component~\cite{Gawthrop2007aa, KarMarRos12}. Thus, for example, the chemical and electrical domains are connected using a transformer with modulus $F$; the Faraday constant. As \BGT are connection permissive, it is up to the modeller, or appropriate software, to use the correct transformer to connect physical domains.

\subsection{Version Controlled Models}
In \BGT\, models are constructed via an API, and hence the entire process is textual. 
This means that models can be version controlled using standard concurrent versioning systems (CVS) such as git, mercurial or svn.
Further, one can essentially assemble and share model libraries in the same way one would package and distribute a python script.
Whilst this is less important in physical domains that have well characterised models, such as electrical engineering, classical mechanics or hydraulics, it is crucial in areas such as systems biology where processes and parameters are not as well understood. It is of critical importance for these fields to have provenance with respect to models, parameters and the relationship of those to experimental protocols and data.
Bond graphs have been shown to be useful in these areas~\cite{Gawthrop2019, Pan2018aa, Pan2018ab} and while the application of \BGT\ here is a subject of future work, it is clear that the utility of integrating efficient and well established `off-the-shelf' version tracking cannot be understated.

\subsection{Sustainable Software Practices}\label{sec:sussoft}
The development of \BGT\ employs many sustainable software techniques which benefit both developers and users.
\BGT\ is available on the python package index (PyPI) and requires python 3.6, or 3.7, the simulation tools also require julia 0.6.4. 
The latest version of \BGT\ can be installed using the console command \texttt{pip3 install BondGraphTools} and requires only the python and julia binaries to be pre-installed.
Further installation guides, tutorial, and code documentation is hosted on readthedocs\footnote{\url{http://bondgraphtools.readthedocs.io}} which is updated automatically as new versions are released.
Source code is accessible online\footnote{\url{https://github.com/BondGraphTools}} and is distributed under the permissive, open-source Apache 2.0 license. 
Semantic versioning via git tags is used to keep track of releases in perpetuity and
Vagrant archives (entirely self-contained Linux images with the software pre-loaded) are available on the Systems Biology Laboratory\footnote{\url{https://systemsbiologylaboratory.org.au/}} website for posterity to capture both the library and dependencies' state at the time of publication to ensure reproducibility~\cite{Dan2018}.

Development of \BGT\ proceeds within an agile iterative and adaptative methodology, as opposed to using a sequential plan-build-test process, which ensures that usable software is prioritised and 
time is not wasted on planning features that users are less interested in. Instead, requirements and design emerge out of the development process and user experiences, with punctuated breaks for refactoring to simplify the codebase.
Test-driven development (TDD) is employed to ensure quality, continuity and stability with automated integration/testing and code metrics provided via Travis.ci\footnote{\url{https://www.travis-ci.com}} and CodeClimate\footnote{\url{www.codeclimate.com}} respectively.
With TDD, unit tests for new software features are specified and implemented before development begins. A feature is then complete once it passes the corresponding unit tests (in addition to the existing tests) prompting integration of the new code into the trunk of code base. TDD is eminently suited to API or library development as developing a battery of tests for common use cases is an integral part of the development process ensuring that existing functionality does not break as new features are added and encouraging developers to think hard about the library interface.
At the time of writing, test coverage is around 80\%, which is respectable in the context of an active, iterative  development process.
 
The feature development and bug status can be tracked using the GitHub issue interface. Users are encourage to log bug reports about any issues they might have.
\subsection{Discussion} \label{sec:conc}

Specialist tools are required to build and analyse complex physical systems and these often take the form of monolithic computer aided design applications, or extensions to mathematical software.
Here we have presented, \BGT; an open source python library for systems modelling which enables the model building process to be scripted, allowing for automation previously unavailable. 
Further, \BGT\ uses best-practice software engineering techniques to increase the sustainability and longevity of the software.
We have demonstrated how \BGT\ can contribute to the modeller's workflow by considering a number of cross domain models in the existing literature.
\BGT\ continues to be actively developed and used across a variety of problem domains, particularly in systems biology, bio-electrics and electrical engineering.
Although basic symbolic model order reduction exists within \BGT, an important future research direction involves developing and implementing scalable symbolic order reduction algorithms for nonlinear systems.
Reduced order equations are a desired output as they are a simpler representation of the system dynamics, giving greater insight into how parameters effect emergent states and speeding up numerical simulations.
This is particularly important for systems with multiple time scales, occurring commonly in biology and engineering, which can pose significant challenges to numerical solvers.
\BGT\ offers an promising platform on which to develop both exact and approximate symbolic reduction algorithms which would be a useful in engineering, scientific computing and computational mathematics.
\section*{Acknowledgements}
This research was in part conducted and funded by the Australian Research Council Centre of Excellence in Convergent Bio-Nano Science and Technology (project number CE140100036).
M.P. acknowledges the financial support provided by an Australian Government Research Training Program Scholarship.


\end{document}